\documentclass[preprintnumbers,amsmath,amssymbm,preprint]{revtex4}
\usepackage{epsfig}
\usepackage{graphicx}

\begin{document}
\title{Kerr-Newman black holes with stationary charged scalar clouds}
\author{Shahar Hod}
\affiliation{The Ruppin Academic Center, Emeq Hefer 40250, Israel}
\affiliation{ } \affiliation{The Hadassah Institute, Jerusalem
91010, Israel}
\date{\today}

\begin{abstract}
\ \ \ It is shown that Kerr-Newman black holes can support linear
charged scalar fields in their exterior regions. To that end, we
solve analytically the Klein-Gordon wave equation for a stationary
charged massive scalar field in the background of a near-extremal
Kerr-Newman black hole. In particular, we derive a simple analytical
formula which describes the physical properties of these stationary
bound-state resonances of the charged massive scalar fields in the
Kerr-Newman black-hole spacetime.
\end{abstract}
\bigskip
\maketitle


\section{Introduction}

The `no-hair' conjecture \cite{Whee,Car} has played a central role
in the physics of black holes since its introduction by Wheeler more
than four decades ago \cite{Bek1,Chas,BekVec,Hart,Nun,Hod11}. This
conjecture suggests that all asymptotically flat stationary
black-hole spacetimes belong to one and the same family --- the
Kerr-Newman family of black holes \cite{Chan,Kerr,Newman}. If true,
this conjecture implies that stationary black holes are
characterized by only three externally observable physical
parameters: mass, charge, and angular momentum.

The influential no-hair conjecture asserts, in particular, that
static fields can not be supported in the spacetime region exterior
to the black-hole horizon
\cite{Whee,Car,Bek1,Chas,BekVec,Hart,Nun,Hod11,Noteelec}. It is
therefore expected that external fields which are not associated
with globally conserved charges would eventually be absorbed into
the black hole or be radiated away to infinity. In accord with this
physically motivated expectation, the dynamics of perturbation
fields in a black-hole spacetime is characterized by quasinormal
ringing, damped oscillations which reflect the gradual dissipation
of energy from the black-hole exterior region \cite{QNMs}. At late
times, these exponentially decaying oscillations are followed by
inverse power-law decaying tails \cite{Tails}.

It is worth emphasizing that existing no-hair theorems
\cite{Whee,Car,Bek1,Chas,BekVec,Hart} do not rule out the existence
of non-static composed black-hole-field configurations. In
particular, it has recently \cite{Hodstat} been demonstrated
explicitly that rotating Kerr black holes can support linearized
stationary scalar configurations (scalar ``clouds"
\cite{Notecloud,Barr}) in their exterior regions. Since non-linear
effects tend to stabilize hairy black-hole configurations
\cite{Nun,Hod11}, we conjectured in \cite{Hodstat} the existence of
genuine non-static hairy black-hole-scalar-field configurations,
which are the non-linear counterparts of the linear scalar clouds
discussed in \cite{Hodstat}. The existence of these non-static hairy
black-hole-scalar-field configurations was demonstrated numerically
most recently in the important work by Herdeiro and Radu
\cite{HerRa}

The composed non-static black-hole-scalar-field configurations
\cite{Notebos} studied in \cite{Hodstat,HerRa} owe their existence
to two distinct physical effects which together combine to trap the
fields in the spacetime region exterior to the black-hole horizon:
\newline
(1) The first physical mechanism responsible for the existence of
these non-static black-hole-bosonic-field configurations
\cite{Hodstat,HerRa} is the well-explored phenomenon of superradiant
scattering of bosonic fields in black-hole spacetimes
\cite{Zel,PressTeu1,Ins1,Ins2,Ins3}. In particular, it is well
established (see \cite{Stro,Beksup,Unrsup} and references therein)
that a charged bosonic field of the form $e^{im\phi}e^{-i\omega t}$
interacting with a charged and rotating Kerr-Newman black hole can
extract energy from the hole if the composed system is in the
superradiant regime \cite{Stro,Beksup,Unrsup}
\begin{equation}\label{Eq1}
\omega<\omega_{\text{c}}\ ,
\end{equation}
where the critical frequency $\omega_{\text{c}}$ for superradiant
scattering is given by \cite{Stro,Beksup,Unrsup}
\begin{equation}\label{Eq2}
\omega_{\text{c}}\equiv m\Omega_{\text{H}}+q\Phi_{\text{H}}\  .
\end{equation}
Here \cite{Noteunits}
\begin{equation}\label{Eq3}
\Omega_{\text{H}}={{a}\over{r^2_++a^2}}\ \ \ {\text{and}} \ \ \
\Phi_{\text{H}}={{Qr_+}\over{r^2_++a^2}}
\end{equation}
are the angular velocity and the electric potential of the black
hole, respectively ($M, Q, Ma$, and $r_+$ are respectively the mass,
charge, angular momentum, and horizon-radius of the black hole). The
parameters $m$ and $q$ in (\ref{Eq2}) are the azimuthal harmonic
index and the charge coupling constant of the bosonic field,
respectively.
\newline
(2) The second physical mechanism required for the existence of
composed non-static black-hole-bosonic-field configurations
\cite{Hodstat,HerRa} is provided by the mutual gravitational
attraction between the central black hole and the {\it massive}
bosonic field. It is well known \cite{Ins2} that the mass term $\mu$
[see Eq. (\ref{Eq7}) below] of a massive field effectively acts as a
reflecting wall which prevents low frequency modes with
\begin{equation}\label{Eq4}
\omega^2<\mu^2
\end{equation}
from escaping to spatial infinity.

The main goal of the present study is to generalize the results of
\cite{Hodstat} to the regime of linearized charged massive scalar
fields interacting with charged and rotating Kerr-Newman black
holes. In particular, we shall show that Kerr-Newman black holes can
support stationary charged scalar fields in their exterior regions.
To that end, we shall solve analytically the Klein-Gordon wave
equation for a stationary charged massive scalar field in the
background of a near-extremal Kerr-Newman black hole. In particular,
we shall derive a simple analytical formula which describes the
physical properties of these stationary charged scalar clouds
\cite{Notecloud} in the Kerr-Newman black-hole spacetime.

\section{Description of the system}

The physical system we consider consists of a test charged scalar
field $\Psi$ coupled to a Kerr-Newman black hole of mass $M$,
angular-momentum $Ma$, and electric charge $Q$. In Boyer-Lindquist
coordinates $(t,r,\theta,\phi)$ the spacetime metric is given by
\cite{Chan,Kerr,Newman}
\begin{eqnarray}\label{Eq5}
ds^2=-{{\Delta}\over{\rho^2}}(dt-a\sin^2\theta
d\phi)^2+{{\rho^2}\over{\Delta}}dr^2+\rho^2
d\theta^2+{{\sin^2\theta}\over{\rho^2}}\big[a
dt-(r^2+a^2)d\phi\big]^2\  ,
\end{eqnarray}
where $\Delta\equiv r^2-2Mr+a^2+Q^2$ and $\rho\equiv
r^2+a^2\cos^2\theta$. The black-hole (event and inner) horizons are
located at the zeroes of $\Delta$:
\begin{equation}\label{Eq6}
r_{\pm}=M\pm(M^2-a^2-Q^2)^{1/2}\  .
\end{equation}

The dynamics of a charged massive scalar field $\Psi$ in the
Kerr-Newman spacetime is governed by the Klein-Gordon wave equation
\cite{Teuk,Stro}
\begin{equation}\label{Eq7}
[(\nabla^\nu-iqA^\nu)(\nabla_{\nu}-iqA_{\nu}) -\mu^2]\Psi=0\  ,
\end{equation}
where $\mu$ and $q$ are respectively the mass and charge coupling
constant of the scalar field \cite{Notedim}, and $A_{\nu}$ is the
electromagnetic potential of the charged black hole. Substituting
the decomposition \cite{Notedec}
\begin{equation}\label{Eq8}
\Psi=\sum_{l,m}e^{im\phi}{S_{lm}}(\theta;a\omega){R_{lm}}(r;a,\omega)e^{-i\omega
t}\ ,
\end{equation}
into the Klein-Gordon wave equation (\ref{Eq7}), one finds
\cite{Stro} that the radial function ${R_{lm}}$ and the angular
function ${S_{lm}}$ can be determined from two coupled ordinary
differential equations [see Eqs. (\ref{Eq9}) and (\ref{Eq11}) below]
of the confluent Heun type \cite{Heun,Fiz1,Teuk,Abram,Stro,Hodasy}.

The angular functions $S_{lm}(\theta;a\omega)$ are known as the
spheroidal harmonics. These functions are solutions of the
characteristic angular equation
\cite{Heun,Fiz1,Teuk,Abram,Stro,Hodasy}
\begin{eqnarray}\label{Eq9}
{1\over {\sin\theta}}{{d}\over{\theta}}\Big(\sin\theta {{d
S_{lm}}\over{d\theta}}\Big) +\Big[K_{lm}+a^2(\mu^2-\omega^2)
-a^2(\mu^2-\omega^2)\cos^2\theta-{{m^2}\over{\sin^2\theta}}\Big]S_{lm}=0\
.
\end{eqnarray}
The spheroidal harmonics $S_{lm}(\theta;a\omega)$ determined from
(\ref{Eq9}) should satisfy regularity boundary conditions at the
poles $\theta=0$ and $\theta=\pi$. These boundary conditions pick
out a discrete set of eigenvalues $\{K_{lm}\}$ which are labeled by
the integers $l$ and $m$ with $l\geq m$ \cite{Abram}. In the regime
$a^2(\mu^2-\omega^2)\lesssim m^2$ [see Eqs. (\ref{Eq19}) and
(\ref{Eq30}) below] one may treat the term
$a^2(\omega^2-\mu^2)\cos^2\theta$ in (\ref{Eq9}) as a small
perturbation to the familiar generalized Legendre equation
\cite{Abram}. This yields the perturbation expansion \cite{Abram}
\begin{equation}\label{Eq10}
K_{lm}+a^2(\mu^2-\omega^2)=l(l+1)+\sum_{k=1}^{\infty}c_ka^{2k}(\mu^2-\omega^2)^k\
\end{equation}
for the angular eigenvalues (separation constants) $K_{lm}$. The
expansion coefficients $\{c_k(l,m)\}$ are given in \cite{Abram}.

The radial Teukolsky equation is given by \cite{Teuk,Stro}
\begin{equation}\label{Eq11}
{{d}
\over{dr}}\Big(\Delta{{dR_{lm}}\over{dr}}\Big)+\Big[{{H^2}\over{\Delta}}
+2ma\omega-\mu^2(r^2+a^2)-K_{lm}\Big]R_{lm}=0\ ,
\end{equation}
where
\begin{equation}\label{Eq12}
H\equiv (r^2+a^2)\omega-am-qQr\  .
\end{equation}
Note that the angular eigenvalues $\{{K_{lm}}(a\omega)\}$ couple the
radial Teukolsky equation (\ref{Eq11}) to the angular equation
(\ref{Eq9}) \cite{Notebr}.

In the present paper we shall analyze the bound-state resonances of
the charged massive scalar fields in the Kerr-Newman black-hole
spacetime. These exponentially decaying solutions are characterized
by the asymptotic behavior \cite{Ins2}
\begin{equation}\label{Eq13}
R(r\to\infty)\sim {{1}\over{r}}e^{-\sqrt{\mu^2-\omega^2}r}\  .
\end{equation}
Note that asymptotically decaying bound-states are characterized by
$\omega^2<\mu^2$ [see Eq. (\ref{Eq4})]. In addition, we shall use
the physically motivated boundary condition of purely ingoing waves
(as measured by a comoving observer) at the outer horizon $r=r_+$ of
the black hole \cite{Ins2}:
\begin{equation}\label{Eq14}
R(r\to r_+)\sim e^{-i(\omega-\omega_{c})y}\  ,
\end{equation}
where $\omega_{\text{c}}$ is determined in (\ref{Eq2}) and the
``tortoise" radial coordinate $y$ is defined by
$dy/dr=(r^2+a^2)/\Delta$ \cite{Noteho}.

The boundary conditions (\ref{Eq13}) and (\ref{Eq14}) single out a
discrete family of complex frequencies
$\{\omega(\mu,q,l,m,M,a,Q;n)\}$ \cite{Notenr} which correspond to
the bound-state resonances of the charged massive scalar fields in
the Kerr-Newman black-hole spacetime. The {\it stationary}
bound-state resonances of the charged massive fields, which are the
resonances we shall be interested in in this study, are
characterized by $\Im\omega=0$.

\section{The stationary bound-state resonances of the charged massive fields in the Kerr-Newman black-hole spacetime}

We shall now prove that, a charged scalar wave field with the
critical frequency $\omega=\omega_{\text{c}}$ [the critical
frequency for superradiant scattering, see Eq. (\ref{Eq2})]
corresponds to a stationary bound-state resonance of the charged
massive scalar field in the Kerr-Newman black-hole spacetime. In
particular, in this section we shall derive an analytical formula
for the discrete spectrum of scalar field masses,
$\{\mu(q,l,m,M,a,Q;n)\}$ \cite{Notenr}, which correspond to the
stationary ($\Im\omega=0$) bound-state resonances of the charged
massive scalar fields.

It proves useful to define new dimensionless variables
\cite{Teuk,Stro}
\begin{equation}\label{Eq15}
x\equiv {{r-r_+}\over {r_+}}\ \ ;\ \ \tau\equiv {{r_+-r_-}\over
{r_+}}\ \ ;\ \ k\equiv 2\omega_{\text{c}} r_+-qQ\ ,
\end{equation}
in terms of which the radial Teukolsky equation (\ref{Eq11}) becomes
\begin{equation}\label{Eq16}
x(x+\tau){{d^2R}\over{dx^2}}+(2x+\tau){{dR}\over{dx}}+VR=0\  ,
\end{equation}
where $V\equiv
H^2/r^2_+x(x+\tau)-K+2ma\omega_{\text{c}}-\mu^2[r^2_+(x+1)^2+a^2]$
and $H=r^2_+\omega_{\text{c}}x^2+r_+kx$. We shall henceforth
consider near-extremal black holes with
\begin{equation}\label{Eq17}
\tau\ll1.
\end{equation}

We shall first study the radial equation (\ref{Eq16}) in the region
$x\gg \tau$. In this far region the radial equation (\ref{Eq16}) is
well approximated by
\begin{equation}\label{Eq18}
x^2{{d^2R}\over{dx^2}}+2x{{dR}\over{dx}}+V_{\text{far}}R=0\  ,
\end{equation}
where
$V_{\text{{far}}}=(r_+\omega_{\text{c}}x+k)^2-K+2ma\omega_{\text{c}}-\mu^2[r^2_+(x+1)^2+a^2]$.
Defining the dimensionless quantity
\begin{equation}\label{Eq19}
\epsilon\equiv \sqrt{\mu^2-\omega_{\text{c}}^2}r_+\  ,
\end{equation}
one finds that the solution of Eq. (\ref{Eq18})
is given by
\cite{Morse,Abram}:
\begin{equation}\label{Eq20}
R(x)=C_1\times(2\epsilon)^{{1\over 2}+\beta}x^{-{1\over
2}+\beta}e^{-\epsilon x}M({1\over 2}+\beta-\kappa,1+2\beta,2\epsilon
x)+C_2\times(\beta\to -\beta)\ ,
\end{equation}
where $M(a,b,z)$ is the confluent hypergeometric function
\cite{Abram} and $\{C_1,C_2\}$ are normalization constants to be
determined below. Here
\begin{equation}\label{Eq21}
\beta^2\equiv K+{1\over
4}-2ma\omega_{\text{c}}-k^2+\mu^2(r^2_++a^2)\ \ ;\ \ \kappa\equiv
{{\omega_{\text{c}}
k-\mu^2r_+}\over{\sqrt{\mu^2-\omega^2_{\text{c}}}}}\ .
\end{equation}
The notation $(\beta\to -\beta)$ in (\ref{Eq20}) means ``replace
$\beta$ by $-\beta$ in the preceding term." We shall henceforth
consider the case of real $\beta$ \cite{Notebet}.

We shall next study the radial Teukolsky equation (\ref{Eq16}) in
the region $x\ll 1$. The radial equation in this near-horizon region
is given by Eq. (\ref{Eq16}) with $V\to V_{\text{near}}\equiv
-K+2ma\omega_{\text{c}}-\mu^2(r^2_++a^2)+k^2x/(x+\tau)$. The radial
solution which satisfies the
ingoing boundary condition (\ref{Eq14}) at the black-hole horizon is
given by \cite{Morse,Abram}
\begin{equation}\label{Eq22}
R(x)=\Big({x\over \tau}+1\Big)^{-ik}{_2F_1}({1\over
2}+\beta-ik,{1\over 2}-\beta-ik;1;-x/\tau)\  ,
\end{equation}
where $_2F_1(a,b;c;z)$ is the hypergeometric function \cite{Abram}.

For near-extremal Kerr-Newman black holes with $\tau\ll1$ [see Eq.
(\ref{Eq17})], there is an overlap region $\tau\ll x\ll 1$ in which
the two radial expressions, (\ref{Eq20}) and (\ref{Eq22}), can be
matched. The $x\ll 1$ limit of Eq. (\ref{Eq20}) is given by
\cite{Morse,Abram}
\begin{equation}\label{Eq23}
R\to C_1\times(2\epsilon)^{{1\over 2}+\beta}x^{-{1\over
2}+\beta}+C_2\times(\beta\to -\beta)\  ,
\end{equation}
and the $x\gg \tau$ limit of Eq. (\ref{Eq22}) is given by
\cite{Morse,Abram}
\begin{equation}\label{Eq24}
R\to \tau^{{1\over 2}-\beta}{{\Gamma(2\beta)}\over{\Gamma({1\over
2}+\beta-ik)\Gamma({1\over 2}+\beta+ik)}}x^{-{1\over
2}+\beta}+(\beta\to -\beta)\  .
\end{equation}
Matching the expressions (\ref{Eq23}) and (\ref{Eq24}) in the
overlap region $\tau\ll x\ll 1$, one finds the two normalization
constants $\{C_1,C_2\}$ of the radial function (\ref{Eq20}):
\begin{equation}\label{Eq25}
C_1(\beta)=\tau^{{1\over
2}-\beta}{{\Gamma(2\beta)}\over{\Gamma({1\over
2}+\beta-ik)\Gamma({1\over 2}+\beta+ik)}}(2\epsilon)^{-{1\over
2}-\beta}\ \ \ \ {\text{and}} \ \ \ \ C_2(\beta)=C_1(-\beta)\ .
\end{equation}

The asymptotic $x\to\infty$ limit of the radial function
(\ref{Eq20}) is given by \cite{Morse,Abram}
\begin{eqnarray}\label{Eq26}
R(x\to\infty)&\to&
\Big[C_1\times(2\epsilon)^{\kappa}{{\Gamma(1+2\beta)}\over{\Gamma({1\over
2}+\beta+\kappa)}}x^{-1+\kappa}(-1)^{-{1\over
2}-\beta+\kappa}+C_2\times(\beta\to -\beta)\Big]e^{-\epsilon x}
\nonumber \\&& +
\Big[C_1\times(2\epsilon)^{-\kappa}{{\Gamma(1+2\beta)}\over{\Gamma({1\over
2}+\beta-\kappa)}}x^{-1-\kappa}+C_2\times(\beta\to
-\beta)\Big]e^{\epsilon x}\ .
\end{eqnarray}
The bound-state resonances of the charged massive scalar fields are
characterized by exponentially decaying radial solutions at spatial
infinity [see Eq. (\ref{Eq13})]. This implies that the coefficient
of the growing exponent $e^{\epsilon x}$ in (\ref{Eq26}) must be
zero:
\begin{eqnarray}\label{Eq27}
C_1\times(2\epsilon)^{-\kappa}{{\Gamma(1+2\beta)}\over{\Gamma({1\over
2}+\beta-\kappa)}}x^{-1-\kappa}+C_2\times(\beta\to -\beta)=0\  .
\end{eqnarray}
Substituting (\ref{Eq25}) into (\ref{Eq27}), one finds the
characteristic resonance condition
\begin{equation}\label{Eq28}
{1\over{\Gamma({1\over
2}+\beta-\kappa)}}=\Big[{{\Gamma(-2\beta)}\over{\Gamma(2\beta)}}\Big]^2{{\Gamma({1\over
2}+\beta-ik)\Gamma({1\over 2}+\beta+ik)}\over{\Gamma({1\over
2}-\beta-ik)\Gamma({1\over 2}-\beta+ik)\Gamma({1\over
2}-\beta-\kappa)}}\big(2\epsilon\tau\big)^{2\beta}\
\end{equation}
for the stationary bound-state resonances of the charged massive
scalar fields in the Kerr-Newman black-hole spacetime. Note that the
r.h.s. of the resonance condition (\ref{Eq28}) is of order
$O[(\epsilon\tau)^{2\beta}]\ll 1$ [see Eq. (\ref{Eq17}) and Eq.
(\ref{Eq30}) below]. Hence, one may use the well-known pole
structure of the Gamma functions \cite{Abram} in order to write the
characteristic resonance condition (\ref{Eq28}) in the form
\begin{equation}\label{Eq29}
{1\over 2}+\beta-\kappa=-n+O[(\epsilon\tau)^{2\beta}]\ ,
\end{equation}
where $n=0,1,2,...$ is the resonance parameter.

As we shall now show, the characteristic resonance condition
(\ref{Eq29}) for the stationary bound-state resonances of the
charged massive scalar fields in the Kerr-Newman black-hole
spacetime can be solved {\it analytically} in the regime
\begin{equation}\label{Eq30}
\epsilon\ll1\  .
\end{equation}
Taking cognizance of Eqs. (\ref{Eq10}), (\ref{Eq15}), (\ref{Eq19}),
and (\ref{Eq21}), one finds \cite{Notea,Hodnb}
\begin{equation}\label{Eq31}
\beta=\beta_0+O(\epsilon^2)\ \ \ \text{and}\ \ \
\kappa={{\alpha}\over{\epsilon}}+O(\epsilon)\  ,
\end{equation}
in the regime (\ref{Eq30}), where
\begin{equation}\label{Eq32}
\beta_0\equiv \sqrt{\big(l+{1\over
2}\big)^2-2ma\omega_{\text{c}}-(2\omega_{\text{c}}r_+-qQ)^2+\omega^2_{\text{c}}(r^2_++a^2)}\
\ \ \ ; \ \ \ \ \alpha\equiv
\omega_{\text{c}}r_+(\omega_{\text{c}}r_+-qQ)\  .
\end{equation}
Substituting (\ref{Eq31}) into the resonance condition (\ref{Eq29}),
one finds \cite{Notenes}
\begin{equation}\label{Eq33}
\epsilon(q,l,m,M,a,Q;n)={{\alpha}\over{\beta_0+{1\over 2}+n}}\ .
\end{equation}
Finally, we recall that the field-masses which correspond to the
stationary bound-state resonances of the charged massive scalar
fields are given by [see Eq. (\ref{Eq19})]
\begin{equation}\label{Eq34}
\mu r_+(q,l,m,M,a,Q;n)=\sqrt{(\omega_{\text{c}}r_+)^2+\epsilon^2}\ .
\end{equation}

\section{Weakly bound stationary resonances: some physical examples}

It is worth noting that the assumption $\epsilon\ll1$ [see Eq.
(\ref{Eq30})] corresponds to weakly bound-state resonances
($\omega_{\text{c}}\lesssim\mu$) of the charged massive scalar
fields in the Kerr-Newman black-hole spacetime. As we shall now
show, this assumption is satisfied in several distinct physical
regimes:
\newline
(1) Slowly rotating Kerr-Newman black holes with $a\ll M$. In this
case one finds [see Eqs. (\ref{Eq2}) and (\ref{Eq3})]
\begin{equation}\label{Eq35}
\omega_{\text{c}}r_+=qQ+m{\bar a}-qQ{\bar a}^2+O({\bar a}^3)\  ,
\end{equation}
where
\begin{equation}\label{Eq36}
\bar a\equiv {{a}\over{r_+}}\  .
\end{equation}
Substituting (\ref{Eq35}) into (\ref{Eq32}), one obtains
\begin{equation}\label{Eq37}
\beta_0=\big(l+{{1}\over{2}}\big)[1+O({\bar a})]\ \ \ \ ; \ \ \ \
\alpha=qQm{\bar a}[1+O({\bar a})]\ ,
\end{equation}
which implies [see Eq. (\ref{Eq33})]
\begin{equation}\label{Eq38}
\epsilon={{qQm}\over{l+1+n}}\cdot{\bar a}[1+O({\bar a})]\ll1\ .
\end{equation}
Substituting (\ref{Eq35}) and (\ref{Eq38}) into (\ref{Eq34}), one
finds
\begin{equation}\label{Eq39}
\mu r_+=qQ\Big\{1+{{m}\over{qQ}}\cdot{\bar a}+\Big[{1\over
2}\Big({{m}\over{l+1+n}}\Big)^2-1\Big]{\bar a}^2+O({\bar
a}^3)\Big\}\
\end{equation}
for the field-masses of the stationary bound-state resonances.
\newline
(2) Composed Kerr-Newman-scalar-field configurations in the regime
\cite{NoteqQ1}
\begin{equation}\label{Eq40}
\omega_{\text{c}}r_+\ll1\ \ \ \text{with}\ \ \ qQ=O(1) .
\end{equation}
Substituting (\ref{Eq40}) into (\ref{Eq32}), one finds
\begin{equation}\label{Eq41}
\beta_0=\sqrt{\Big(l+{1\over
2}\Big)^2-(qQ)^2}+O(\omega_{\text{c}}r_+)\ \ \ \; \ \ \ \
\alpha=-qQ\omega_{\text{c}}r_+[1+O(\omega_{\text{c}}r_+)]\ ,
\end{equation}
which implies [see Eq. (\ref{Eq33})]
\begin{equation}\label{Eq42}
\epsilon=-{{2qQ}\over{\sqrt{(2l+1)^2-(2qQ)^2}+1+2n}}\cdot
\omega_{\text{c}}r_+[1+O(\omega_{\text{c}}r_+)]\ll1\ .
\end{equation}
Substituting (\ref{Eq42}) into (\ref{Eq34}), one finds
\begin{equation}\label{Eq43}
\mu r_+=\omega_{\text{c}}r_+
\sqrt{1+{{(2qQ)^2}\over{\big[\sqrt{(2l+1)^2-(2qQ)^2}+1+2n\big]^2}}}[1+O(\omega_{\text{c}}r_+)]\
.
\end{equation}
\newline
(3) Composed Kerr-Newman-scalar-field configurations in the regime
\cite{NoteqQ2}
\begin{equation}\label{Eq44}
\omega_{\text{c}}r_+=qQ+\delta\ \ \ \text{with}\ \ \ \delta\ll1\  .
\end{equation}
Substituting (\ref{Eq44}) into (\ref{Eq32}), one finds
\begin{equation}\label{Eq45}
\beta_0=\sqrt{\Big(l+{1\over 2}\Big)^2-m^2}+O(\delta)\ \ \ ; \ \ \
\alpha=qQ\delta[1+O(\delta)]\ ,
\end{equation}
which implies [see Eq. (\ref{Eq33})]
\begin{equation}\label{Eq46}
\epsilon={{2qQ}\over{\sqrt{(2l+1)^2-4m^2}+1+2n}}\cdot{\delta}[1+O(\delta)]\ll1\
.
\end{equation}
Substituting (\ref{Eq44}) and (\ref{Eq46}) into (\ref{Eq34}), one
finds
\begin{equation}\label{Eq47}
\mu
r_+=qQ\Big\{1+{{1}\over{qQ}}\cdot{\delta}+{{2}\over{\big[\sqrt{(2l+1)^2-4m^2}+1+2n\big]^2}}\cdot\delta^2+O(\delta^3)\Big\}\
.
\end{equation}
\newline
(4) Nearly polar scalar clouds with $l\gg {\text{max}}(m,qQ)$. In
this case one finds [see Eqs. (\ref{Eq32}) and (\ref{Eq33})]
\begin{equation}\label{Eq48}
\epsilon={{\alpha}\over{l}}[1+O(l^{-1})]\ll1\ .
\end{equation}
Substituting (\ref{Eq48}) into (\ref{Eq34}), one finds
\begin{equation}\label{Eq49}
\mu
r_+=\omega_{\text{c}}r_++{{\alpha^2}\over{2\omega_{\text{c}}r_+}}\cdot{{1}\over{l^2}}[1+O(l^{-1})]\
.
\end{equation}
\newline
(5) High overtone modes with $n\gg {\text{max}}(l,qQ)$. In this case
one finds [see Eqs. (\ref{Eq32}) and (\ref{Eq33})]
\begin{equation}\label{Eq50}
\epsilon={{\alpha}\over{n}}[1+O(n^{-1})]\ll1\ .
\end{equation}
Substituting (\ref{Eq50}) into (\ref{Eq34}), one finds
\begin{equation}\label{Eq51}
\mu
r_+=\omega_{\text{c}}r_++{{\alpha^2}\over{2\omega_{\text{c}}r_+}}\cdot{{1}\over{n^2}}[1+O(n^{-1})]\
.
\end{equation}

\section{Summary and discussion}

In summary, we have shown that Kerr-Newman black holes can support
linear charged scalar fields in their exterior regions. To that end,
we have solved analytically the Klein-Gordon wave equation for a
stationary charged massive scalar field in the background of a
near-extremal Kerr-Newman black hole. In particular, we have derived
an analytical formula [see Eqs. (\ref{Eq32})-(\ref{Eq34})] which
determines the discrete spectrum of scalar field masses,
$\{\mu(q,l,m,M,a,Q;n)\}$, which correspond to the stationary
bound-state resonances of the charged massive scalar fields in the
Kerr-Newman black-hole spacetime.

It is worth emphasizing that the stationary charged scalar clouds
studied in this paper owe their existence to the rotation of the
central black hole. In particular, we have shown that there are no
stationary bound-state configurations of the charged scalar field
with $ma=0$ \cite{Notea}. This conclusion is in accord with the
results presented in \cite{Hodnb,HerRa} for static charged
Reissner-Nordstr\"om black holes \cite{Carexm}.

Finally, we would like to note that, it would be physically
interesting to generalize  our {\it analytical} results for the
linear charged scalar clouds to the non-linear regime of a genuine
charged scalar hair. Such a generalization would probably require a
fully non-linear {\it numerical} \cite{HerRa} analysis of the
combined Einstein-Maxwell-Klein-Gordon equations.

\bigskip
\noindent
{\bf ACKNOWLEDGMENTS}
\bigskip

This research is supported by the Carmel Science Foundation. I thank
Yael Oren, Arbel M. Ongo and Ayelet B. Lata for stimulating
discussions.



\begin{thebibliography}{99}

\bibitem{Whee} R. Ruffini and J. A. Wheeler, Phys. Today {\bf 24}, 30
(1971).

\bibitem{Car} B. Carter, in {\it Black Holes}, Proceedings of 1972 Session of Ecole d'ete de Physique Theorique,
edited by C. De Witt and B. S. De Witt (Gordon and Breach, New York,
1973).

\bibitem{Bek1} J. D. Bekenstein, Phys. Rev. D {\bf 7}, 2333 (1973);
J. D. Bekenstein, Phys. Today {\bf 33}, 24 (1980); J. D. Bekenstein,
Phys. Rev. D {\bf 51}, R6608 (1995); A. E. Mayo and J. D.
Bekenstein, Phys. Rev. D {\bf 54}, 5059 (1996).

\bibitem{Chas} J. E. Chase, Commun. Math. Phys. {\bf 19}, 276 (1970); J. D. Bekenstein,
Phys. Rev. Lett. {\bf 28}, 452 (1972); C. Teitelboim, Lett. Nuovo
Cimento {\bf 3}, 326 (1972); I. Pena and D. Sudarsky, Class. Quant.
Grav. {\bf 14}, 3131 (1997).

\bibitem{BekVec} J. D. Bekenstein, Phys. Rev. D {\bf 5}, 1239 (1972); {\bf 5}, 2403 (1972);
M. Heusler, J. Math. Phys. {\bf 33}, 3497 (1992); D. Sudarsky,
Class. Quantum Grav. {\bf 12}, 579 (1995).

\bibitem{Hart} J. Hartle, Phys. Rev. D {\bf 3}, 2938 (1971); C. Teitelboim, Lett.
Nuovo Cimento {\bf 3}, 397 (1972).

\bibitem{Nun} D. N\'u\~nez, H. Quevedo, and D. Sudarsky, Phys. Rev. Lett. {\bf 76}, 571 (1996).

\bibitem{Hod11} S. Hod, Phys. Rev. D {\bf 84}, 124030 (2011) [arXiv:1112.3286].

\bibitem{Chan} S. Chandrasekhar, {\it The Mathematical Theory of Black
Holes}, (Oxford University Press, New York, 1983).

\bibitem{Kerr} R. P. Kerr, Phys. Rev. Lett. {\bf 11}, 237 (1963).

\bibitem{Newman} E. T. Newman, R. Couch, K. Chinnapared, A. Exton,
A. Prakash, et. al., J. Math. Phys. {\bf 6}, 918 (1965).

\bibitem{Noteelec} One obvious exception to this statement is provided by the black-hole electric field.
However, this field is associated with a globally conserved charge.

\bibitem{QNMs} H. P. Nollert, Class. Quantum Grav. {\bf 16}, R159 (1999);
E. Berti, V. Cardoso and A. O. Starinets, Class. Quant. Grav. {\bf
26}, 163001 (2009); E. W. Leaver, Proc. R. Soc. A {\bf 402}, 285
(1985); B. Mashhoon, Phys. Rev. D {\bf 31}, 290 (1985); L. E. Simone
and C. M. Will, Class. Quantum Grav. {\bf 9}, 963 (1992); H. P.
Nollert, Phys. Rev. D {\bf 47}, 5253 (1993); S. Hod, Phys. Rev.
Lett. {\bf 81}, 4293 (1998) [arXiv:gr-qc/9812002]; G. T. Horowitz
and V. E. Hubeny, Phys. Rev. D {\bf 62}, 024027 (2000); K.
Glampedakis and N. Andersson, Phys. Rev. D {\bf 64}, 104021 (2001);
S. Hod, Phys. Rev. D {\bf 67}, 081501 (2003) [arXiv:gr-qc/0301122];
S. Hod and U. Keshet, Class. Quant. Grav. {\bf 22}, L71 (2005)
[arXiv:gr-qc/0505112]; S. Hod and U. Keshet, Phys.Rev. D {\bf 73},
024003 (2006) [arXiv:hep-th/0506214]; S. Hod, Class. Quant. Grav.
{\bf 23}, L23 (2006) [arXiv:gr-qc/0511047]; U. Keshet and S. Hod,
Phys. Rev. D {\bf 76}, R061501 (2007) [arXiv:0705.1179]; S. Hod,
Phys. Rev. D {\bf 75}, 064013 (2007) [arXiv:gr-qc/0611004]; S. Hod,
Class. and Quant. Grav. {\bf 24}, 4235 (2007) [arXiv:0705.2306]; A.
Gruzinov, arXiv:gr-qc/0705.1725; A. Pesci, Class. Quantum Grav. {\bf
24}, 6219 (2007); S. Hod, Phys. Rev. D {\bf 78}, 084035 (2008)
[arXiv:0811.3806]; S. Hod, Phys. Lett. B {\bf 666} 483 (2008)
[arXiv:0810.5419]; S. Hod, Phys. Rev. D {\bf 80}, 064004 (2009)
[arXiv:0909.0314]; V. Cardoso, A. S. Miranda, E. Berti, H. Witek,
and V. T. Zanchin, Phys. Rev. D {\bf 79}, 064016 (2009); S. Hod,
Phys. Lett. A {\bf 374}, 2901 (2010) [arXiv:1006.4439]; R. A.
Konoplya and A. Zhidenko, Rev. Mod. Phys. {\bf 83}, 793 (2011); S.
Hod, Phys. Rev. D. {\bf 84}, 044046 (2011) [arXiv:1109.4080]; Y.
D\'ecanini, A. Folacci, and B. Raffaelli, Phys. Rev. D {\bf 84},
084035 (2011); S. Hod, Phys. Lett. B {\bf 710}, 349 (2012)
[arXiv:1205.5087]; R. A. Konoplya, A. Zhidenko, Phys. Rev. D {\bf
88}, 024054 (2013); S. Hod, Phys. Lett. B {\bf 715}, 348 (2012)
[arXiv:1207.5282]; H. Yang, A. Zimmerman, A. Zenginoglu, F. Zhang,
E. Berti, and Yanbei Chen, Phys. Rev. D {\bf 88}, 044047 (2013); S.
Hod, Phys. Rev. D {\bf 88}, 084018 (2013) [arXiv:1311.3007].

\bibitem{Tails} R. H. Price, Phys. Rev. D {\bf 5}, 2419 (1972);
C. Gundlach, R. H. Price, and J. Pullin, Phys. Rev. D {\bf 49}, 883
(1994); J. Bic\'ak, Gen. Relativ. Gravitation {\bf 3}, 331 (1972);
E. S. C. Ching, P. T. Leung, W. M. Suen, and K. Young, Phys. Rev.
Lett. {\bf 74}, 2414 (1995); E. S. C. Ching, P. T. Leung, W. M.
Suen, and K. Young, Phys. Rev. D {\bf 52}, 2118 (1995); S. Hod and
T. Piran, Phys. Rev. D {\bf 58}, 024017 (1998)
[arXiv:gr-qc/9712041]; S. Hod and T. Piran, Phys. Rev. D {\bf 58},
024018 (1998) [arXiv:gr-qc/9801001]; S. Hod and T. Piran, Phys. Rev.
D {\bf 58}, 044018 (1998) [arXiv:gr-qc/9801059]; S. Hod and T.
Piran, Phys. Rev. D {\bf 58}, 024019 (1998) [arXiv:gr-qc/9801060];
S. Hod, Phys. Rev. D {\bf 58}, 104022 (1998) [arXiv:gr-qc/9811032];
S. Hod, Phys. Rev. D {\bf 61}, 024033 (2000) [arXiv:gr-qc/9902072];
S. Hod, Phys. Rev. D {\bf 61}, 064018 (2000) [arXiv:gr-qc/9902073];
L. Barack, Phys. Rev. D {\bf 61}, 024026 (2000); S. Hod, Phys. Rev.
Lett. {\bf 84}, 10 (2000) [arXiv:gr-qc/9907096]; S. Hod, Phys. Rev.
D {\bf 60}, 104053 (1999) [arXiv:gr-qc/9907044]; S. Hod, Class.
Quant. Grav. {\bf 26}, 028001 (2009) [arXiv:0902.0237]; S. Hod,
Class. Quant. Grav. {\bf 18}, 1311 (2001) [arXiv:gr-qc/0008001]; S.
Hod, Phys. Rev. D {\bf 66}, 024001 (2002) [arXiv:gr-qc/0201017]; R.
J. Gleiser, R. H. Price, and J. Pullin, Class. Quant. Grav. {\bf
25}, 072001 (2008); M. Tiglio, L. E. Kidder, and S. A. Teukolsky,
Class. Quant. Grav. {\bf 25}, 105022 (2008); R. Moderski and M.
Rogatko, Phys. Rev. D {\bf 77}, 124007 (2008); X. He and J. Jing,
Nucl. Phys.B {\bf 755}, 313 (2006); H. Koyama and A. Tomimatsu,
Phys. Rev. D {\bf 65}, 084031 (2002); B. Wang, C. Molina, and E.
Abdalla, Phys. Rev. D {\bf 63}, 084001 (2001); A. Zenginoglu and M.
Tiglio, Phys. Rev.D {\bf 80}, 024044 (2009); A. J. Amsel, G.  T.
Horowitz, D. Marolf, and M. M. Roberts, J. High Energy Phys.
0909:044 (2009); S. Hod, Class. and Quant. Grav. {\bf 30}, 237002
(2013) [arXiv:1402.4819]; S. Hod, Jour. of High Energy Phys. {\bf
1309} (2013) 056 [arXiv:1310.4247].

\bibitem{Hodstat} S. Hod, Phys. Rev. D {\bf 86}, 104026 (2012)
[arXiv:1211.3202]; S. Hod, The Euro. Phys. Journal C {\bf 73}, 2378
(2013) [arXiv:1311.5298].

\bibitem{Notecloud} We use here the terminology of \cite{Barr} to reflect
the fact that these exterior stationary matter configurations are
made of test (linearized) scalar fields.

\bibitem{Barr} J. Barranco, A. Bernal, J. C. Degollado, A. Diez-Tejedor,
M. Megevand, et al., Phys. Rev. Lett. {\bf 109}, 081102 (2012);
Phys. Rev. D {\bf 84}, 083008 (2011).

\bibitem{HerRa} C. A. R. Herdeiro and E. Radu, Phys. Rev. Lett.
(2014) [arXiv:1403.2757].

\bibitem{Notebos} Other bosonic fields are also expected to form
rotating hairy black-hole configurations, see also \cite{HerRa}.

\bibitem{Zel} Ya. B. Zel'dovich, Pis`ma Zh. Eksp. Teor. Fiz. {\bf
14}, 270 (1971) [JETP Lett. {\bf 14}, 180 (1971)]; Zh. Eksp. Teor.
Fiz. {\bf 62}, 2076 (1972) [Sov. Phys. JETP {\bf 35}, 1085 (1972)];
A. V. Vilenkin, Phys. Lett. B {\bf 78}, 301 (1978).

\bibitem{PressTeu1} W. H. Press and S. A. Teukolsky, Nature {\bf
238}, 211 (1972); W. H. Press and S. A. Teukolsky, Astrophys. J.
{\bf 185}, 649 (1973).

\bibitem{Ins1} V. Cardoso, O. J. C. Dias, J. P. S. Lemos and S.
Yoshida, Phys. Rev. D {\bf 70}, 044039 (2004) [Erratum-ibid. D {\bf
70}, 049903 (2004)]; J. C. Degollado, C. A. R. Herdeiro, and H. F.
R\'unarsson, Phys. Rev. D {\bf 88}, 063003 (2013); J. C. Degollado
and C. A. R. Herdeiro, Phys. Rev. D {\bf 89}, 063005 (2014); S. Hod,
Phys. Rev. D {\bf 88}, 064055 (2013) [arXiv:1310.6101]; R. Li,
arXiv:1404.6309; S. Hod, Phys. Rev. D {\bf 88}, 124007 (2013)
[arXiv:1405.1045].

\bibitem{Ins2} T. Damour, N. Deruelle and R. Ruffini, Lett. Nuovo Cimento {\bf
15}, 257 (1976); T. M. Zouros and D. M. Eardley, Annals of physics
{\bf 118}, 139 (1979); S. Detweiler, Phys. Rev. D {\bf 22}, 2323
(1980); H. Furuhashi and Y. Nambu, Prog. Theor. Phys. {\bf 112}, 983
(2004); V. Cardoso and J. P. S. Lemos, Phys. Lett. B {\bf 621}, 219
(2005); V. Cardoso and S. Yoshida, JHEP 0507:009 (2005); S. R.
Dolan, Phys. Rev. D {\bf 76}, 084001 (2007); S. Hod and O. Hod,
Phys. Rev. D {\bf 81}, Rapid communication 061502 (2010)
[arXiv:0910.0734]; H. R. Beyer, J. Math. Phys. {\bf 52}, 102502
(2011); S. Hod, Phys. Lett. B {\bf 708}, 320 (2012)
[arXiv:1205.1872]; R. Brito, V. Cardoso, and P. Pani, Phys. Rev. D
{\bf 88}, 023514 (2013); S. R. Dolan, Phys. Rev. D {\bf 87}, 124026
(2013); H. Witek, V. Cardoso, A. Ishibashi, and U. Sperhake, Phys.
Rev. D {\bf 87}, 043513 (2013); V. Cardoso, Gen. Relativ. and
Gravit. {\bf 45}, 2079 (2013); H. Okawa, H. Witek, and V. Cardoso,
Phys. Rev. D {\bf 89}, 104032 (2014).

\bibitem{Ins3} V. Cardoso and O. J. C. Dias, Phys. Rev. D {\bf 70}, 084011 (2004);
O. J. C. Dias, G. T. Horowitz and J. E. Santos, JHEP {\bf 1107}
(2011) 115; O. J. C. Dias, P. Figueras, S. Minwalla, P. Mitra, R.
Monteiro and J. E. Santos, JHEP {\bf 1208} (2012) 117; O. J. C. Dias
and J. E. Santos, JHEP {\bf 1310} (2013) 156; V. Cardoso, O. J. C.
Dias, G. S. Hartnett, L. Lehner and J. E. Santos, arXiv:1312.5323.

\bibitem{Stro} T. Hartman, W. Song, and A. Strominger, JHEP 1003:118 (2010).

\bibitem{Beksup} J. D. Bekenstein, Phys. Rev. D {\bf 7}, 949 (1973).

\bibitem{Unrsup} M. Richartz, S. Weinfurtner, A. J. Penner and W. G.
Unruh, Phys. Rev. D {\bf 80}, 124016 (2009).

\bibitem{Noteunits} We use natural units in which
$G=c=\hbar=1$.

\bibitem{Teuk} S. A. Teukolsky, Phys. Rev. Lett. {\bf 29}, 1114 (1972);
S. A. Teukolsky, Astrophys. J. {\bf 185}, 635 (1973).

\bibitem{Notedim} Note that the field parameters $q$ and $\mu$ stand for $q/\hbar$ and $\mu/\hbar$,
respectively. Hence, they have the dimensions of $($length$)^{-1}$.

\bibitem{Notedec} Here $\omega$ is the conserved frequency of the wave field, and $(l,m)$ are
respectively the spheroidal harmonic index and the azimuthal
harmonic index of the mode [see Eq. (\ref{Eq9}) below].

\bibitem{Heun} A. Ronveaux, {\it Heun's differential equations}.
(Oxford University Press, Oxford, UK, 1995); C. Flammer, {\it
Spheroidal Wave Functions} (Stanford University Press, Stanford,
1957).

\bibitem{Fiz1} P. P. Fiziev, e-print arXiv:0902.1277; R. S. Borissov and P. P. Fiziev, e-print arXiv:0903.3617;
P. P. Fiziev, Phys. Rev. D {\bf 80}, 124001 (2009); P. P. Fiziev,
Class. Quant. Grav. {\bf 27}, 135001 (2010).

\bibitem{Abram} M. Abramowitz and I. A. Stegun, {\it Handbook of
Mathematical Functions} (Dover Publications, New York, 1970).

\bibitem{Hodasy} S. Hod, Phys. Rev. Lett. {\bf 100}, 121101 (2008) [arXiv:0805.3873];
S. Hod, Phys. Lett. B {\bf 717}, 462 (2012) [arXiv:1304.0529]; S.
Hod, Phys. Rev. D {\bf 87}, 064017 (2013) [arXiv:1304.4683].

\bibitem{Notebr} We shall henceforth omit the harmonic indexes $l$ and $m$ for brevity.

\bibitem{Noteho} Note that $r\to r_+$ corresponds to $y\to -\infty$.

\bibitem{Notenr} Here $n=0,1,2,...$ is the resonance parameter.

\bibitem{Morse} P. M. Morse and H. Feshbach, {\it Methods of
Theoretical Physics} (McGraw-Hill, New York, 1953).

\bibitem{Notebet} One can choose $\beta>0$ without loss of generality.

\bibitem{Notea} We consider the case of rotating ($a\neq0$)
Kerr-Newman black holes. For $a=0$ (a non-rotating charged
Reissner-Nordstr\"om black hole) one finds $\alpha=0$ and
$\kappa=-\epsilon$, in which case the resonance condition
(\ref{Eq29}) has no solutions (note that $\beta_0+1/2>0$ and
$-\kappa>0$ in this case). One therefore concludes that non-rotating
charged Reissner-Nordstr\"om black holes cannot support stationary
scalar clouds. This conclusion is in accord with the results
presented in \cite{Hodnb} for Reissner-Nordstr\"om black holes.

\bibitem{Hodnb} S. Hod, Phys. Lett. B {\bf 713}, 505 (2012); S. Hod, Phys. Lett. B {\bf 718}, 1489 (2013)
[arXiv:1304.6474].

\bibitem{Notenes} Taking cognizance of Eqs. (\ref{Eq29}) and (\ref{Eq31}), one realizes
that $\kappa>0$ (or equivalently, $\alpha>0$) is a necessary
condition for the existence of bound-state resonances.

\bibitem{NoteqQ1} Note that the regime (\ref{Eq40}) corresponds to $qQ\simeq -m{\bar
a}$, see Eqs. (\ref{Eq2}) and (\ref{Eq3}).

\bibitem{NoteqQ2} Note that the regime (\ref{Eq44}) corresponds to $qQ=m{\bar a}^{-1}-(1+{\bar
a}^{-2})\cdot\delta$, see Eqs. (\ref{Eq2}) and (\ref{Eq3}).

\bibitem{Carexm} For the case of marginally bound ($\omega=\mu$) states of a charged scalar
field in the extremal Reissner-Nordstr\"om black-hole spacetime, see
J. C. Degollado and C. A. R. Herdeiro, Gen. Rel. Grav. {\bf 45},
2483 (2013).

\end{thebibliography}
\end{document}